\documentclass[]{tMPH2e}

\usepackage{multirow}
\usepackage{array}
\usepackage{hhline}
\usepackage{upgreek}

\begin{document}

\doi{10.1080/0026897YYxxxxxxxx}
\issn{13623028}
\issnp{00268976}
\jvol{00}
\jnum{28} \jyear{2016} \jmonth{ june}

\articletype{ARTICLE}

\title{High-resolution photoelectron-spectroscopic investigation of the H$_2$O$^+$ cation in its $\tilde{\mathrm{A}}^+$ electronic state}

\author{C. Lauzin$^{a,b}$, B. Gans$^{a,c}$ and F. Merkt$^{a,\ast}$\thanks{$^\ast$Corresponding author. Email: merkt@phys.chem.ethz.ch
\vspace{6pt}}\\ \vspace{6pt} $^{a}${\em{Physical Chemistry Laboratory, ETH Z\"urich, 8093 Z\"urich, Switzerland.\\
$^b$ Institute of Condensed Matter and Nanosciences, Universit\'{e} Catholique de Louvain, 1348 Louvain-la-Neuve, Belgium.\\
$^c$ Universit\'{e} Paris-Sud, Institut des Sciences Mol\'eculaires d'Orsay (ISMO), UMR 8214, Orsay, F-91405, France.}}}

\maketitle

\begin{abstract}
The photoelectron spectrum of water has been recorded in the vicinity of the $\tilde{\mathrm{A}}^+ \leftarrow \tilde{\mathrm{X}}$ transition between 112 000 and 116 000 cm$^{-1}$ (13.89-14.38 eV). The high-resolution allowed the observation of the rotational structure of several bands. Rotational assignments of the transitions involving the $\Pi(080)$, $\Sigma(070)$ and $\Pi(060)$ vibronic states of the $\tilde{\mathrm{A}}^+$ electronic state are deduced from previous studies of the $\tilde{\mathrm{A}}^+ - \tilde{\mathrm{X}}^+$ band system of H$_2$O$^+$ (Lew, \textit{Can. J. Phys.} \textbf{54}, 2028 (1976) and Huet \textit{et al.} \textit{J. Chem. Phys.} \textbf{107}, 5645 (1997)) and photoionization selection rules. The transition to the $\Sigma(030)$ vibronic state is tentatively assigned.

\end{abstract}

\section{Introduction}
The lowest two electronic states of the H$_2$O$^+$ cation ($\tilde{\mathrm{X}}^+\,^2\mathrm{B}_1$ and $\tilde{\mathrm{A}}^+\,^2\mathrm{A}_1$) form a degenerate pair of $^2\Pi_{\mathrm{u}}$ symmetry in the linear configuration (point group D$_{\infty \mathrm{h}}$). This degeneracy is lifted by the Renner-Teller effect as soon as the molecule bends, the coupling being mediated by the bending mode $\nu^+_2$. This pair of states and the vibronic interaction between them have been the object of extensive work, both experimental \cite{Lew1973, Karlsson1974, Lew1976, Dixon1976, Reutt1986, Tsuji1988, Das1991, Huet1992, Dressler1995, Huet1997, Truong2009, Ford2010, Lauzin2015} and theoretical \cite{Jungen1980,Jungen1980b, Jungen1980c,Brommer1992,Meyer2010}.
The most extensive set of high-resolution spectroscopic information on the $\tilde{\mathrm{X}}^+$ and $\tilde{\mathrm{A}}^+$ states of H$_2$O$^+$ have been obtained from spectra of the $\tilde{\mathrm{A}}^+ - \tilde{\mathrm{X}}^+$ band system in the visible and near infrared ranges of the electromagnetic spectrum \cite{Lew1973, Lew1976, Huet1997, Gan2004}. Because of the considerable difference in bending angles between the equilibrium structures of the $\tilde{\mathrm{X}}^+$ (bending angle of 109$^ \circ$) and the $\tilde{\mathrm{A}}^+$ state (linear or quasilinear structure),
the lowest vibrational levels of $\tilde{\mathrm{A}}^+$ ($v^{\textrm{lin}}_2<6$) have not been observed so far in studies of the $\tilde{\mathrm{A}}^+ - \tilde{\mathrm{X}}^+$ band system, and detailed analyzes of the rotational structure of the $\tilde{\mathrm{A}}^+$ state have been limited to $\tilde{\mathrm{A}}^+$ vibrational levels with $v^{\textrm{lin}}_2\geq 6$. 

Information on the $\tilde{\mathrm{A}}^+$ and $\tilde{\mathrm{X}}^+$ states of H$_2$O$^+$ has also been obtained from photoelectron spectra  \cite{Dixon1976, Reutt1986,Truong2009, Ford2010}. These spectra consist of a very short vibrational progression in the case of the $\tilde{\mathrm{X}}^+ - \tilde{\mathrm{X}}$ ionizing transition and of a very long vibrational progression in the bending mode $\nu_2^+$ in the case of the $\tilde{\mathrm{A}}^+ - \tilde{\mathrm{X}}$ ionizing transition, reflecting the similar equilibrium structure of the  $\tilde{\mathrm{X}}$ and  $\tilde{\mathrm{X}^+}$ states and the very different equilibrium bending angles of the $\tilde{\mathrm{X}}$ and $\tilde{\mathrm{A}}^+$ states. In earlier work, the weakness of the transitions to the lowest bending levels of the $\tilde{\mathrm{A}}^+$ state and the possible contributions of hot bands made the assignment of the bending progression, in particular of the value of $\nu_2^+$, very difficult \cite{Dixon1976, Reutt1986,Truong2009, Ford2010}. Definitive assignments of the vibrational structure of the $\tilde{\mathrm{A}}^+ - \tilde{\mathrm{X}}$ ionizing transition only became possible by considering the results of the high-resolution spectroscopic measurements of the $\tilde{\mathrm{A}}^+ - \tilde{\mathrm{X}}^+$ band system mentioned above, which themselves remained ambiguous until theoretical predictions became accurate enough to distinguish between neighbouring bending levels of the $\tilde{\mathrm{A}}^+$ state \cite{Jungen1980, Jungen1980b, Brommer1992,Dressler1995}. The latest photoelectron spectra \cite{Reutt1986, Truong2009, Ford2010} reveal a rapidly decreasing intensity down to the adiabatic ionization threshold (see Fig. ~\ref{fig2}a)-c)). Ford \textit{et al.} \cite{Ford2010} even observed structured rotational contours and interpreted them in the realm of the Renner-Teller effect and photoionization selection rules.

The calculations of Brommer \textit{et al.} \cite{Brommer1992} provided a complete list of rovibrational level energies for the $\tilde{\mathrm{A}}^+$ state, indicating a linear equilibrium structure. Although  the agreement between calculated and measured level energies can now be considered excellent, experimental measurements of the $\tilde{\mathrm{A}}^+$ level structure below $v^{\textrm{lin}}_2=6$ are lacking with which the predictions of Brommer \textit{et al.} could be compared.

The present investigation aims at studying the rotational structure of the photoelectron spectrum of the $\tilde{\mathrm{A}}^+ - \tilde{\mathrm{X}}$ transition by PFI-ZEKE photoelectron spectroscopy.  Rotational intensity distributions provide information on the parity of the photoelectron partial waves and help characterizing the threshold ionization dynamics. They may also help detecting a possible deviation of the $\tilde{\mathrm{A}}^+$ state from a linear equilibrium structure.

\section{Experimental section}
\label{YO}
The experimental setup has been described previously \cite{rupper04a}. It consists of a broadly-tunable vacuum-ultraviolet (VUV) laser source coupled with a pulsed-field-ionization zero-kinetic-energy (PFI-ZEKE) photoelectron spectrometer and a pulsed molecular beam. The pulsed VUV laser radiation was generated by two Nd:YAG-pumped dye lasers (repetition rate 25 Hz, pulse duration $\simeq$ 5 ns) by resonant sum-frequency ($\tilde{\nu}_{\mathrm{VUV}}=2\tilde{\nu}_1+\tilde{\nu}_2$) mixing  in a pulsed atomic beam of krypton using the $(4\mathrm{p})^6 \rightarrow (4\mathrm{p})^5 \; 5\mathrm{p}[1/2]_0 $ two-photon resonance of krypton at $2\tilde{\nu}_1$ $=$ 94092.8673 cm$^{-1}$. The VUV laser bandwidth was 0.5 cm$^{-1}$. The VUV wavenumber was scanned by scanning the wavenumber $\tilde{\nu}_2$ of the second dye laser, which was calibrated by recording optogalvanic spectra of Ne and Ar, resulting in an accuracy of 0.5 cm$^{-1}$ for the VUV wavenumber. 

The gas sample was a mixture of He and H$_2$O in a pressure ratio of $\simeq$100:2 obtained by bubbling pure argon gas through a stainless-steel cylinder filled with distilled H$_2$O. The mixture was introduced into the experimental chamber using a pulsed valve operated at a stagnation pressure of 1.5 bar and synchronized with the laser pulses. The supersonic expansion passed through a skimmer before intersecting the VUV laser beam at right angles on the axis of a cylindrical stack made of 5 extraction plates used to apply sequences of two pulsed electric fields. These sequences were applied 2~$\mathrm{\mu}$s after photoexcitation and consisted in a positive discrimination pulse (+ 100 mV/cm) to eliminate prompt electrons and electrons produced by field ionization of the highest Rydberg states, followed by a negative extraction pulse to field ionize Rydberg states with principal quantum number \textit{n} around 200 and extract the electrons toward a microchannel plate detector. 

The PFI-ZEKE photoelectron spectra were recorded by monitoring the electron signal produced by the extraction pulse as a function of the VUV wavenumber. Depending on the amplitude of the discrimination and extraction field pulses, a resolution in the range between 1 and 2 cm$^{-1}$ resulted. This was sufficient to resolve the rotational structure of the photoelectron spectrum. To determine ionization energies, the shift induced by the electric fields, between 1.2 and 1.7 cm$^{-1}$ depending on the amplitudes of the discrimination and extraction pulses, was corrected for as described in Ref.~\cite{Hollenstein2001}.

Using Kr as a nonlinear gas to generate VUV radiation in the range between 111 500 and 116 000 cm$^{-1}$,  where transitions from the ground state of H$_2$O to the lowest rovibrational levels of the $\tilde{\mathrm{A}}^+$ state of H$_2$O$^+$ are located, offers the advantage of very high VUV pulse energies (up to 10$^{11}$ photons/pulse). These high pulse energies are the result of an enhancement of the four-wave mixing process through resonances with Rydberg states of Kr converging to the $^2\mathrm{P}_{3/2}$ and $^2\mathrm{P}_{1/2}$ ionization threshold. They turned out to be crucial to obtain high-resolution PFI-ZEKE photoelectron spectra of the very weak transitions to low vibrational levels of the $\tilde{\mathrm{A}}^+$ state. These resonances in the sum-frequency mixing and the resulting variations of the phase-matching conditions also led to a rapid variation of the intensity with the VUV wavenumber (see Fig.~\ref{fig2}e) below) and to several regions, marked by areas shaded in grey in Fig. \ref{fig2}, where the VUV intensity was too low to observe any signal.


\section{Results and discussion}

\subsection{State labels and photoionization selection rules }

The $\tilde{\mathrm{X}}\ ^1\mathrm{A}_1$ ground electronic state of H$_2$O has a bent equilibrium structure of C$_{2\mathrm{v}}$ point-group symmetry with a bending angle of $\simeq 109 ^\circ$ and thus three vibrational modes, the symmetric stretching mode ($\nu_1$, $\mathrm{a}_1$ symmetry), the bending mode ($\nu_2$, $\mathrm{a}_1$ symmetry) and the asymmetric stretching mode ($\nu_3$, $\mathrm{a}_2$  symmetry). The rovibrational levels of the ground state are labeled ($v_1$,$v_2$,$v_3$)$N_{K_{\mathrm{a}} K_{\mathrm{c}}}$, where $v_i$ (\textit{i}=1-3) are the vibrational quantum numbers of the modes, $N$ is the rotational-angular-momentum quantum number, and $K_{\mathrm{a}}$ and $K_{\mathrm{c}}$ are the usual asymmetric-top projection quantum numbers. At the low temperature of the H$_2$O sample in the supersonic beam, only the rovibrational ground states (000) 0$_{00}$ and (000) 1$_{01}$ of para (spin-statistical weight of 1) and ortho (spin-statistical weight of 3) H$_2$O are significantly populated. At 10 K, the Boltzman factors of the (000) 1$_{11}$ and (000) 1$_{10}$ levels only amount to 5.10$^{-3}$ and 3.7 10$^{-2}$, respectively, if one assumes that ortho-para conversion does not take place under the expansion conditions of our supersonic beam. Rotational levels with $N\geq2$ were not observed at all. 

\begin{figure}[!h]
 \centering
 \begin{minipage}[c]{.48\linewidth}
 \includegraphics[scale=0.4]{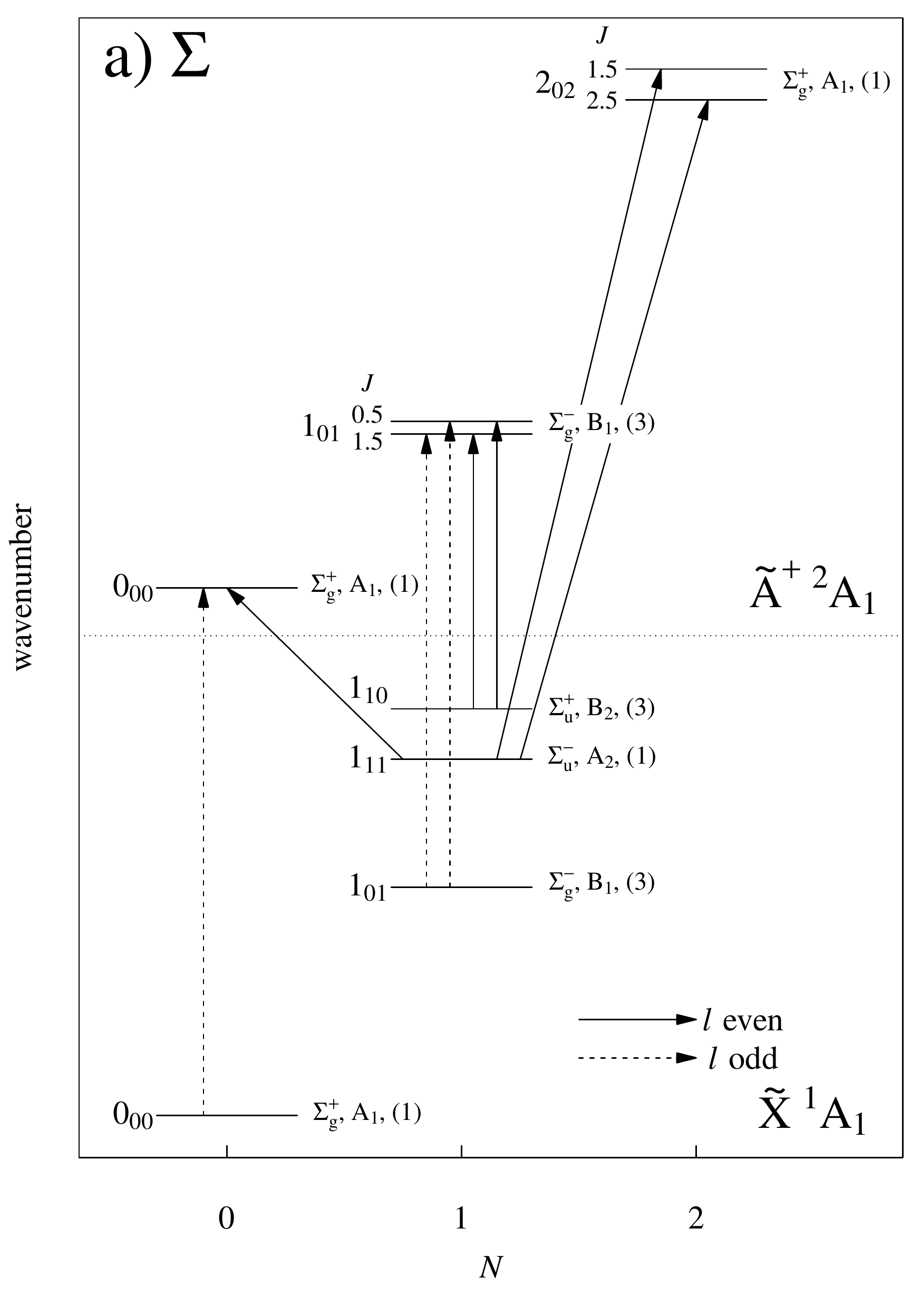}
 \end{minipage} \hfill
  \begin{minipage}[c]{.48\linewidth}
  \includegraphics[scale=0.4]{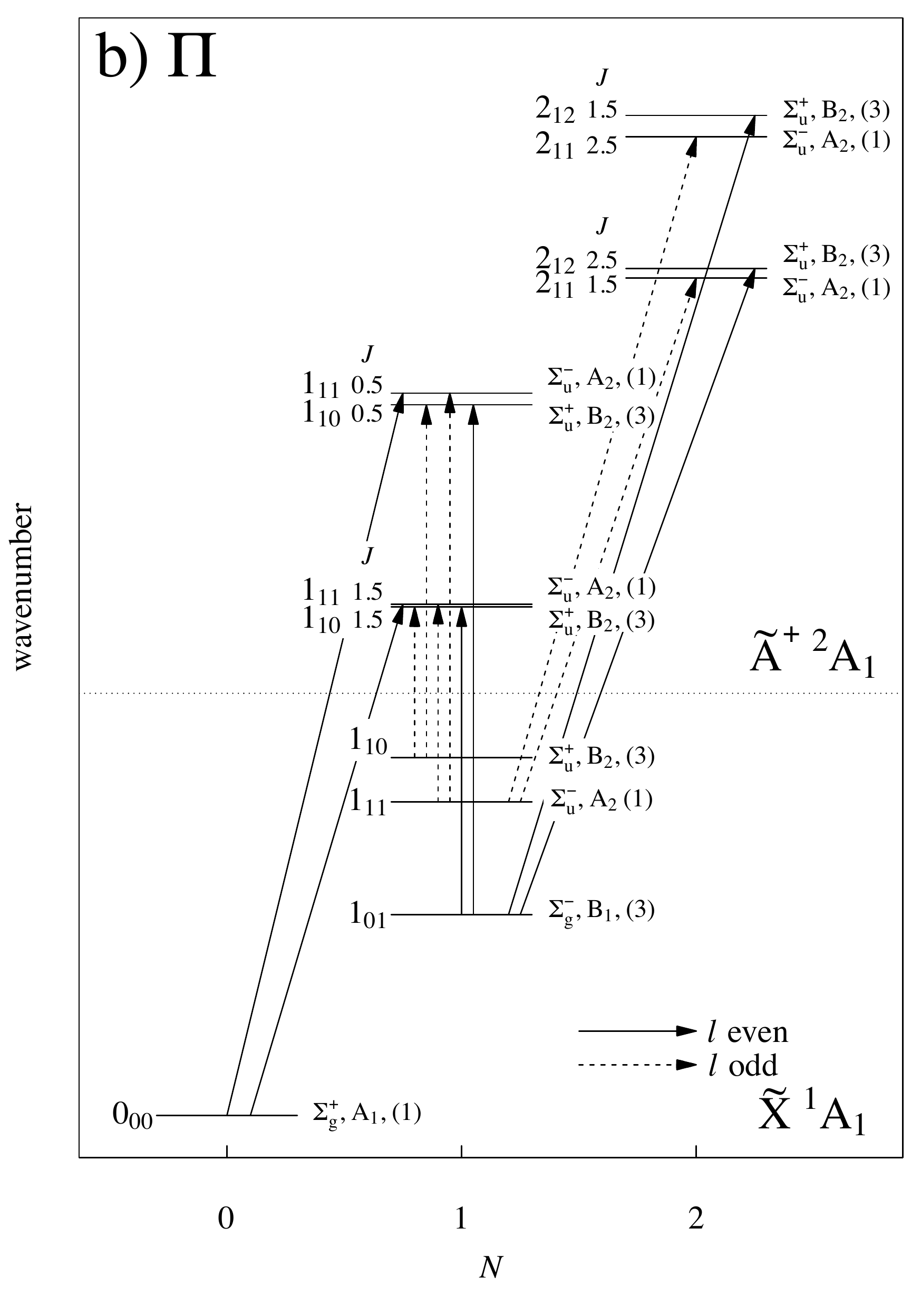}
 \end{minipage}
 \caption{Allowed transitions for the H$_2$O$^+$ $\tilde{\mathrm{A}}^+$ $\Sigma (070)$ $\leftarrow$  H$_2$O $\tilde{\textrm{X}} (000)$ band for an outgoing electron with even (full lines) or odd (dashed lines) values of \textit{l}. Only the $\Delta N=0, \pm$1 and $\Delta K_{\mathrm{c}}=0, \pm$1  transitions are illustrated for the sake of clarity. The number in parentheses after the symmetry labels represent the spin-statistical weights, \textit{i.e.}, 3 for ortho H$_2$O$^{(+)}$ and 1 for para H$_2$O$^{(+)}$.}
 \label{fig1}
\end{figure}

The level structure, with symmetry labels in the C$_{2\mathrm{v}}(\mathrm{M})$ and D$_{\infty \mathrm{h}}(\mathrm{M})$ molecular symmetry groups, is depicted schematically in the lower part of Fig \ref{fig1}.  
The symmetry of the rotational levels is unambiguously set by the even/odd nature of the asymmetric-top quantum numbers $K_{\mathrm{a}}$ and $K_{\mathrm{c}}$, as summarized in Table \ref{parity}, which also provides the correlation between symmetry labels in the C$_{2\mathrm{v}}$(M) and D$_{\infty \mathrm{h}}$(M) groups. Unlike the $\tilde{\mathrm{X}}$  $^1\mathrm{A}_1$ of H$_2$O and the $\tilde{\mathrm{X}}^+$ $^2\mathrm{B}_1$ ground states of H$_2$O$^+$, the $\tilde{\mathrm{A}}^+$ $^2\mathrm{A}_1$ state of H$_2$O$^+$ has a linear (or quasilinear) structure, $^2\Pi_{\mathrm{u}}$ electronic symmetry, and four vibrational modes, the symmetric stretching mode ($\nu_1$, $\sigma^+_{\mathrm{g}}$ point-group symmetry), the doubly degenerate bending mode ($\nu_2$, $\pi_{\mathrm{u}}$ point-group symmetry) and the asymmetric stretching mode ($\nu_3$, $\sigma_{\mathrm{u}}^+$ point-group symmetry). A further quantum number, $K=\mid l_{\mathrm{v}} \pm \varLambda \mid$, is used to label the vibronic states, where $\varLambda$ is the quantum number associated with the projection of the electronic orbital angular momentum onto the molecular axis ($\varLambda=1$ for the $\tilde{\mathrm{A}}^+$ state), and $l_{\mathrm{v}}$ is the vibrational angular-momentum quantum number arising from the doubly degenerate bending mode \cite{Herzberg, Bunker}. The rotational levels can be classified in Hund's case (b) nomenclature with the rotation quantum number $N$ and the total angular momentum quantum number $J$ ($J=N\pm S$ with $S= \frac{1}{2}$, where $S$ is the total-electron-spin quantum number). The vibronic levels of the $\tilde{\mathrm{A}}^+$ state can thus be labelled as $K(v_1,v_2,v_3)$ , where $K$ is indicated by a capital Greek letter ($K=0,1,2,..$ $\rightarrow$ $\Sigma, \Pi, \Delta, ...)$. When treating the Renner-Teller effect, the levels can be labelled in notations adequate for a bent or a linear structure and the two notations can be interchanged using Table \ref{parity} and the correlations $\mid K\mid \leftrightarrow \mid K_{\mathrm{a}}\mid$ and $v_2^{\textrm{lin}}\leftrightarrow (2v^{\textrm{bent}}_2+K_{\mathrm{a}}+1$) \cite{Jungen1980, Jungen1980b, Huet1992, Huet1997}. To conform with earlier works \cite{Lew1976, Huet1997}, we label the rotational levels of the $\tilde{\mathrm{X}}$ state of H$_2$O and $\tilde{\mathrm{A}}^+$ state of H$_2$O$^+$ using the asymmetric-top notation $N_{K_{\mathrm{a}}K_{\mathrm{c}}}$ and $N^+_{K_{\mathrm{a}}^+K_{\mathrm{c}}^+}$, so that the full label for a rovibronic transition is $K(v_1,v^{\textrm{lin}}_2,v_3)N^+_{K^+_{\mathrm{a}}K^+_{\mathrm{c}}}\leftarrow (000)N_{K_{\mathrm{a}}K_{\mathrm{c}}}$.

The general rovibronic photoionization selection rules for molecular photoionization are \cite{Signorell1997}:
\begin{equation}
\Gamma_\mathrm{rve}(\mathrm{ion})\otimes \Gamma_\mathrm{rve}(\mathrm{neutral}) \supset \Gamma^* \text{ for $\it{l}$ even}
\label{eq1}
\end{equation}

and 
\begin{equation}
\Gamma_\mathrm{rve}(\mathrm{ion})\otimes \Gamma_\mathrm{rve}(\mathrm{neutral}) \supset \Gamma^{(\mathrm{s})} \text{ for $\it{l}$ odd,}\label{eq2}
\end{equation}
where $\it{l}$ is the angular momentum quantum number of the photoelectron partial wave, $\Gamma_\mathrm{rve}(\mathrm{ion})$ and $\Gamma_\mathrm{rve}(\mathrm{neutral})$ are the representations of the rovibronic states of the neutral molecule and the molecular ion, respectively. These representations need to be in the same molecular-symmetry group, \textit{i.e.}, the C$_{2\mathrm{v}}(\mathrm{M})$ or the D$_{\infty \mathrm{h}}(\mathrm{M})$ molecular-symmetry group. In C$_{2\mathrm{v}}(\mathrm{M})$, one obtains the following selection rules:

\begin{subequations}
\begin{equation}
\mathrm{A}_1 \longleftrightarrow \mathrm{A}_1, \mathrm{A}_2 \longleftrightarrow \mathrm{A}_2,\mathrm{B}_1 \longleftrightarrow \mathrm{B}_1,\mathrm{B}_2 \longleftrightarrow \mathrm{B}_2
\label{eq3a}
\end{equation}
or equivalently
\begin{equation}
\Delta K_\mathrm{a}=\text{even} , \Delta K_\mathrm{c}=\text{even}
\label{eq3b}
\end{equation}
\end{subequations}
 for ${l}$ odd, and
\begin{subequations}
\begin{equation}
\mathrm{A}_1 \longleftrightarrow \mathrm{A}_2,\mathrm{B}_1 \longleftrightarrow \mathrm{B}_2
\label{eq4a}
\end{equation}
or equivalently
\begin{equation}
\Delta K_\mathrm{a}=\text{odd} , \Delta K_\mathrm{c}=\text{odd}
\label{eq4b}
\end{equation}
\end{subequations}
for ${l}$ even.

\begin{table}[!h]
\caption{Symmetry species of $N_{K_\mathrm{a} K_\mathrm{c}}$ and $N^+_{K^+_\mathrm{a} K^+_\mathrm{c}}$ rotational levels of the H$_2$O $\tilde{\mathrm{X}}$ and H$_2$O$^+$ $\tilde{\mathrm{A}}^+$ electronic state in the C$_{2 \mathrm{v}}(\mathrm{M})$ and D$_{\infty \mathrm{h}}(\mathrm{M})$ molecular-symmetry groups. In the first column, $e$ and $o$ indicate an even or an odd value of $K_{\mathrm{a}}$ or $K_{\mathrm{c}}$}
\centering
\begin{tabular}{ccc}
\hline
$K_\mathrm{a}\ K_\mathrm{c}$ & $\Gamma_\mathrm{rot}$ in C$_{2 \mathrm{v}}$(M) & $\Gamma_\mathrm{rot}$ in D$_{\infty \mathrm{h}}(\mathrm{M})$  \\
\hline
 $ee$ &  $\mathrm{A}_1$ & $\Sigma _\mathrm{g}^+$ \\
$eo$ & $\mathrm{B}_1$ & $\Sigma _\mathrm{g}^-$ \\
$oe$ & $\mathrm{B}_2$ & $\Sigma _\mathrm{u}^+$ \\
$oo$ & $\mathrm{A}_2$ & $\Sigma _\mathrm{u}^-$ \\
\hline
\end{tabular}
\label{parity}
\end{table} 
Panels a) and b) of Fig.~\ref{fig1} illustrate the allowed  rovibronic photoionizing transitions from the $\tilde{\mathrm{X}}$(000) state of H$_2$O to $\Sigma$ and $\Pi$ vibronic levels of the $\tilde{\mathrm{A}}^+$ state of H$_2$O, respectively. Full and dashed lines correspond to even and odd values of $l$. The molecular orbitals out of which the electrons are ejected in the $\tilde{\mathrm{X}}^+\leftarrow \tilde{\mathrm{X}}$ and $\tilde{\mathrm{A}}^+\leftarrow \tilde{\mathrm{X}}$ photoionizing transitions are p orbitals centred on the oxygen atom with the orbital axes perpendicular to the a axis. The use of the orbital-ionization approximation discussed in Ref. \cite{Willitsch2005} indicates a preference for emission of $l$=0, 2 ($i.\ e.$, $l$ ${\textrm{even}}$, see Eqs. ~\ref{eq4a} and \ref{eq4b}) photoelectron partial waves and for photoionizing transitions with $\mid \Delta K_{\mathrm{a}} \mid=1$ and $\Delta N=0,\pm 1$. Indeed the electron hole generated upon ionization has p character and a nodal plane containing ($\tilde{\mathrm{X}}^+\leftarrow \tilde{\mathrm{X}}$) or located very close ($\tilde{\mathrm{A}}^+\leftarrow \tilde{\mathrm{X}}$) to the a axis. These transitions are indicated as full arrows in Fig. \ref{fig1}. One may also observe weaker transitions associated with odd-$l$ partial waves and these are represented as dashed arrows in Fig. \ref{fig1}.

\subsection{Photoelectron spectra and their assignments}
\begin{figure}[h]
 \centering
 \includegraphics[width=0.8\columnwidth]{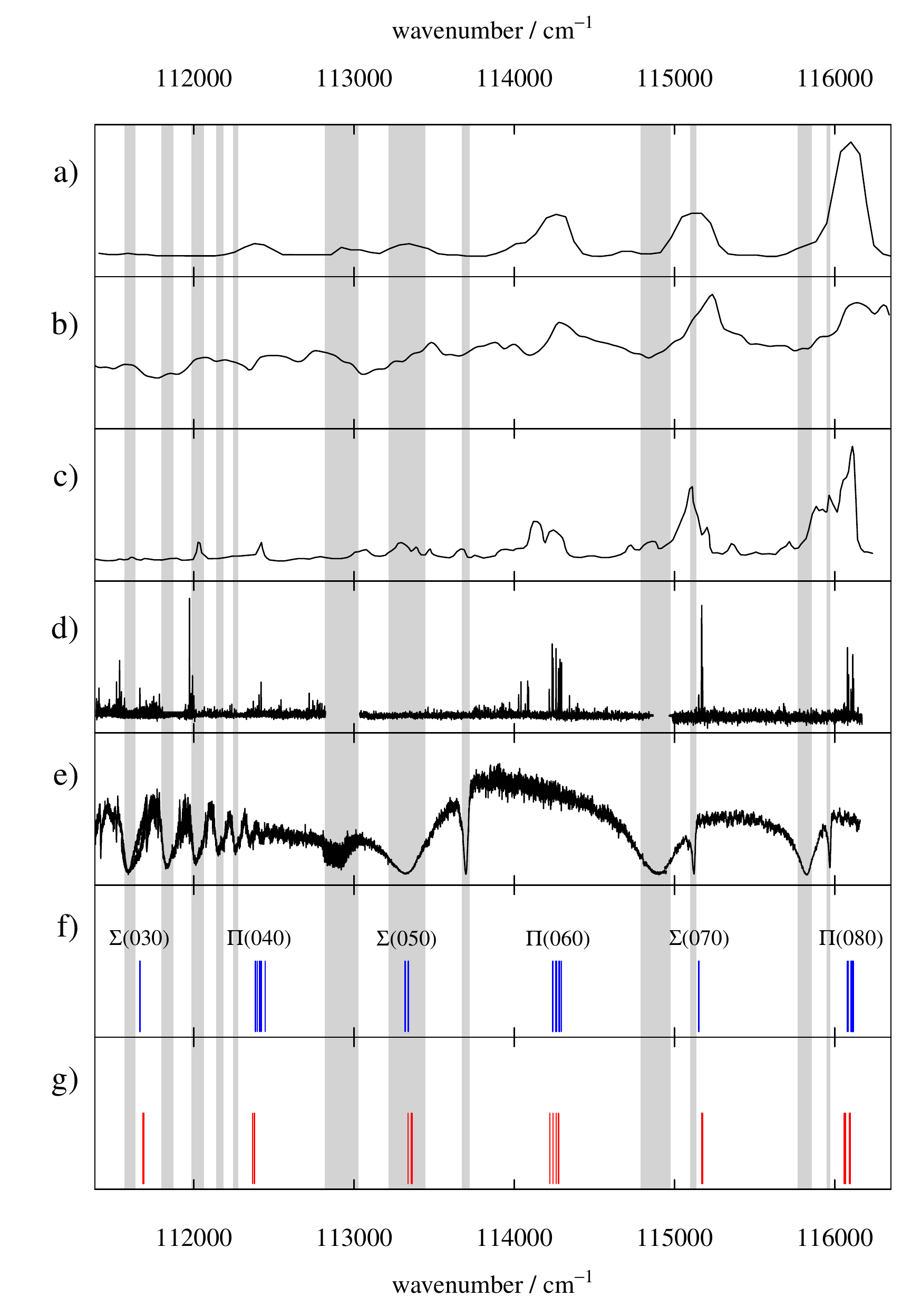}
 \caption{Photoelectron spectra from a) Reutt \textit{et al.} \cite{Reutt1986}, b) Truong \textit{et al.}  \cite{Truong2009}, c) Ford \textit{et al.} \cite {Ford2010}, and d) this work. Panel e) depicts the VUV intensity distribution used to record the spectrum in panel d). Panels f) and g) display the rovibronic photoionization transitions determined from \textit{ab initio} predictions of Brommer \textit{et al.}  \cite{Brommer1992} for photoelectron partial waves with even and odd-\textit{l} values, respectively and restricted to $\Sigma$ and $\Pi$ vibronic final states. The grey-shaded areas mark regions where the VUV intensity was too weak to record a PFI-ZEKE photoelectron spectrum of sufficient quality. The data presented in panels a) - c) were extracted from the original publications in a process that occasionally reduced the high quality of the published spectra, especially for the spectrum presented in panel c).}
 \label{fig2}
\end{figure}

Fig. \ref{fig2} compares our PFI-ZEKE photoelectron spectrum of the $\tilde{\mathrm{A}}^+ -\tilde{\mathrm{X}}$ photoelectron band of H$_2$O (panel d)) with the high-resolution photoelectron spectra reported earlier by Reutt \textit{et al.} \cite{Reutt1986} (panel a)), Truong \textit{et al.} \cite{Truong2009} (panel b)) and Ford \textit{et al.} \cite{Ford2010} (panel c)). As explained in Section 2, the strong variations of the VUV laser intensitities with wavenumber (see panel e)) prevented us from recording spectra of sufficient quality in the regions shaded in grey in Fig. \ref{fig2} and also strongly affected the relative intensities of the observed transitions. Reutt \textit{et al.} \cite{Reutt1986} performed their measurement with a cold supersonic beam and a He I radiation source, achieving a resolution of 11 meV. The spectrum reported by Truong \textit{et al.} \cite{Truong2009} is a threshold photoelectron spectrum of a room-temperature sample recorded at a resolution of about 8 meV. Ford \textit{et al.} \cite{Ford2010} also employed a He I radiation source but used an effusive (room-temperature) beam to reduce Doppler broadening and achieved a resolution of 2 meV. In the energy range depicted in Fig. \ref{fig2}, all four photoelectron spectra reveal a long vibrational progression corresponding to (0,$v^{\textrm{lin}}_2$,0) bending levels of the $\tilde{\mathrm{A}}^+$ state of H$_2$O$^+$ with $v^{\textrm{lin}}_2$ ranging from 3 to 8. The agreement between these spectra recorded under very different experimental conditions can be regarded as very good.

\begin{figure}[h]
 \centering
 \includegraphics[width=1.0\columnwidth]{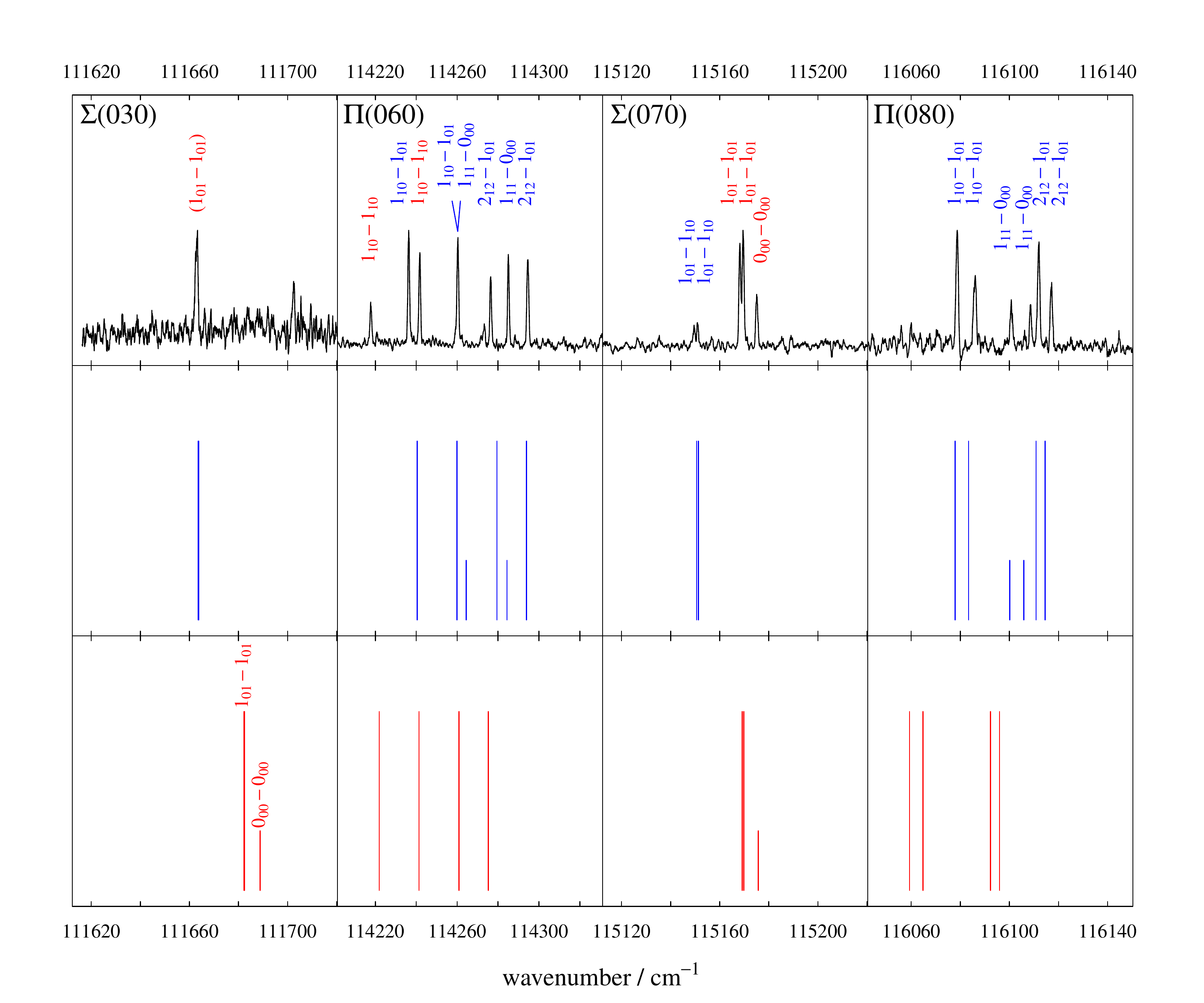}
 \caption{Top row, from left to right: Rotational structure of the PFI-ZEKE photoelectron spectrum of the $\tilde{\mathrm{A}}^+\ (0,v_2^{\textrm{lin}},0)\leftarrow \tilde{\mathrm{X}}(000)$ bands with $v_2^{\textrm{lin}}=3$, 6, 7 and 8. Middle row: Stick spectrum of transitions determined from the calculated term values of Brommer \textit{et al.} \cite{Brommer1992} for even-\textit{l} photoelectron partial waves. Bottom row: As middle row, but for odd-\textit{l} photoelectron partial waves.}
 \label{fig3}
\end{figure}

The two lowest panels of Fig. \ref{fig2} indicate the positions of the rovibronic transitions we determined from $\tilde{\mathrm{A}}^+$ levels calculated by Brommer \textit{et al.}  \cite{Brommer1992} and which we have separated into two groups, one corresponding to even-$l$ (panel f)) and one to odd-$l$ (panel g)) photoelectron partial waves according to Eqs. (4) and (3), respectively. The good agreement between the calculated transition wavenumbers and the bands observed in the experimental spectra provides unambiguous assignments of the transitions to the $\Pi(060)$, $\Sigma(070)$, and $\Pi(080)$ vibronic levels and plausible assignments of the transitions to the $\Sigma(030)$ and $\Pi(040)$ levels. Unfortunately, a hole in the VUV intensity profile around 113 400 cm$^{-1}$ prevented us from obtaining information in the region where the transition to the $\Sigma (050)$ vibronic level is expected. The photoelectron spectra also reveal structures, e.g., around 112 000 and 114 050 cm$^{-1}$ in panel d), where no transition to $\Sigma$ and $\Pi$ vibronic states are expected from the calculations. 
\begin{table}[!h]
\centering
\caption{ Term values (in cm$^{-1}$) of the observed rovibrational levels of the $\tilde{\mathrm{A}}^+$ state of H$_2$O$^+$. The positions of the F$_1$, $J=N+\frac{1}{2}$  and  F$_2$, $J=N-\frac{1}{2}$ levels always appears as the first and second entries, respectively.}
\begin{tabular}{c c c c}
 $N^+ _{K_\mathrm{a} K_\mathrm{c}}$ & This work  & Previous experimental studies & \textit{ab initio} predictions from Ref. \cite{Brommer1992}\   \\
\hline
\multicolumn{4}{c}{ } \\
\multicolumn{4}{c}{$\Pi(080)$} \\
\multirow{2}{*}{$1_{11}$} &14334.4(11)  & 14334.06 \cite{Lew1976} & 14333.85 \\
& 14342.3(11) & 14341.81 \cite{Lew1976} & 14339.56\\
\multicolumn{4}{c}{ } \\
\multirow{2}{*}{$1_{10}$} & 14336.2(11) & 14335.60 \cite{Lew1976} & 14335.39 \\
& 14343.2(11) & 14342.93 \cite{Lew1976} & 14340.88 \\
\multicolumn{4}{c}{ } \\
\multirow{2}{*}{$2_{12}$} & 14369.5(11) & 14368.78 \cite{Lew1976} & 14368.37\\
& 14374.6(11) & 14373.97 \cite{Lew1976} & 14372.05\\
\multicolumn{4}{c}{ } \\
\multicolumn{4}{c}{$\Sigma(070)$} \\
 $0_{00}$ & 13408.8(11) & 13409.263 \cite{Huet1997} & 13409.45\\
 \multicolumn{4}{c}{ } \\
\multirow{2}{*}{$1_{01}$} & 13427.0(11) & 13426.611 \cite{Huet1997} & 13427.35 \\
 & 13425.6(11) & 13425.293 \cite{Huet1997} & 13426.65\\
 \multicolumn{4}{c}{ } \\
\multicolumn{4}{c}{$\Pi(060)$} \\
\multirow{2}{*}{$1_{10}$} & 12493.9(11) & 12493.236 \cite{Huet1997} & 12498.00 \\
&  12517.8(11) & 12517.168 \cite{Huet1997} & 12517.45 \\
\multicolumn{4}{c}{ } \\
\multirow{2}{*}{$1_{11}$} & 12494.0(11) & 12493.544 \cite{Huet1997} & 12498.15  \\
 & 12518.7(11) & 12518.564 \cite{Huet1997} & 12518.09 \\
 \multicolumn{4}{c}{ } \\
\multirow{2}{*}{$2_{12}$} & 12533.8(11) & 12533.331 \cite{Huet1997} & 12536.94\\
& 12551.9(11) & 12551.454 \cite{Huet1997} & 12551.34\\
\multicolumn{4}{c}{ } \\
\multicolumn{4}{c}{$\Sigma (030)$} \\
\multirow{2}{*}{$1_{01}$} & \multirow{2}{*}{9920.5(11)} & & 9939.89 \\
& & & 9939.68\\
\hline
\end{tabular}
\label{positions}
\end{table} 
The rotational structure of the PFI-ZEKE photoelectron spectrum in the regions of the $\Sigma (030)$, $\Pi(060)$, and $\Pi(080)$ is presented in more detail in Fig. \ref{fig3}, where it is compared with stick spectra generated from the \textit{ab initio} calculations of Brommer \textit{et al.} for even-$l$ (middle row) and odd-$l$ (bottom row) partial waves. The intensities of the sticks in these spectra only reflect the spin-statistical weights of ortho (3) and para (1) H$_2$O. Term values (in cm$^{-1}$) of the observed rovibrational levels of the $\tilde{\mathrm{A}}^+$ state of H$_2$O$^+$ are listed in Table \ref{positions} where they are compared with earlier results.

The spectrum of the transition to the $\Pi (080)$ level consists of the six lines expected for even-\textit{l} photoelectron partial waves (see Eqs. (\ref{eq3a}), (\ref{eq3b}) and Fig. \ref{fig1}) from the ground states of para (0$_{00}$) and ortho (1$_{01}$) H$_2$O. The line positions closely correspond to the $\tilde{\mathrm{A}}^+\ \Pi(080)$ rotational level structure determined from the $\tilde{\mathrm{A}}^+ -\tilde{\mathrm{X}}^+$ band system by Lew \cite{Lew1976} (see Table \ref{positions}) and are also in good agreement with the $\tilde{\mathrm{A}}^+\ \Pi(080)$ level positions calculated by Brommer \textit{et al.} \cite{Brommer1992}. The term values listed in Table \ref{positions} were determined by adding the ground-state rotational energies to the field-corrected transition wavenumbers observed in the PFI-ZEKE photoelectron spectra and subtracting the adiabatic ionization energy of the $\tilde{\mathrm{X}}^+-\tilde{\mathrm{X}}$ transition (\textit{i.e.}, 101 766.3(10) cm$^{-1}$\cite{Lauzin2015}).

The same transitions are also observed in the spectrum involving the $\Pi(060)$ vibronic state, which, however, contains in addition the two components of the $1_{10}-1_{10}$ transition that are observed despite the reduced Boltzman factor of the $1_{10}$ ground-state level (see Fig. \ref{fig1} and discussion in Section 3.1), and the fact that these transitions correspond to an odd-$\textit{l}$ photoelectron partial wave. The $\tilde{\mathrm{A}}^+\ \Pi (060)$ rotational term values we derive from our spectrum agree with the term values determined from the $\tilde{\mathrm{A}}^+-\tilde{\mathrm{X}}^+$ band system by Huet \textit{et al.} \cite{Huet1997} within the experimental uncertainty, which leaves no doubt concerning the assignment. 

Transitions corresponding to odd-\textit{l} photoelectron partial waves are also observed in the spectrum of the $\Sigma (070)$ state, where they are even found to be by far the strongest ones. This behaviour can be explained by the fact that the most strongly populated ground-state levels (0$_{00}$ for para H$_2$O and 1$_{01}$ for ortho H$_2$O) both have $K_{\mathrm{a}}=0$ and that $K$ is zero for an $\tilde{\mathrm{A}}^+$ state of $\Sigma$ vibronic symmetry. Consequently, there can be no $\Delta K_{\mathrm{a}}=\pm 1$ transitions from these two levels in the spectrum of the $\Sigma (070)$ state. The only transitions observed from the $\tilde{\mathrm{X}}$ $^1A_1$ $0_{00}$ and $1_{01}$ levels are those corresponding to odd-\textit{l} photoelectron partial waves, \textit{i.e.}, $0_{00}$-$0_{00}$ and $1_{01}$-$1_{01}$. The last pair of transitions observed in this spectrum is attributed to the 1$_{01}$-1$_{10}$ transition and is accompanied by emission of even-\textit{l} partial waves according to Eqs. (\ref{eq3a}) and (\ref{eq3b}). These transitions are weak because of the low Boltzmann factor of the $1_{10}$ level. The positions of the rotational levels of the $\tilde{\mathrm{A}}^+\ \Sigma (070)$ state agree with those determined by Huet \textit{et al.} \cite{Huet1997} from the analysis of the $\tilde{\mathrm{A}}^+-\tilde{\mathrm{X}}^+$ band system and also with the positions calculated by Brommer \textit{et al.} \cite{Brommer1992} (see Table \ref{positions}). 

Only one line was observed in the spectral region where the $\tilde{\mathrm{A}}^+\ \Sigma (030)-\tilde{\mathrm{X}} (000)$ band is expected. From the intensity distribution observed in the spectrum of the  $\tilde{\mathrm{A}}^+\   \Sigma (070)$ state and the arguments presented to explain the dominance of the $1_{01}$-$1_{01}$ transition in this spectrum, we tentatively assign the only line of the $\tilde{\mathrm{A}}^+\ \Sigma(030)$-$\tilde{\mathrm{X}}(000)$ band to the $1_{01}$-$1_{01}$ transition. The alternative assignment would be the 1$_{01}$ - 1$_{10}$ transition,  which, though allowed, originates from the 1$_{10}$ ground state level which has a low Boltzmann factor.  The rotational structure of the spectrum of the $\tilde{\mathrm{A}}^+$ $\Pi(040)-\tilde{\mathrm{X}}(000)$ band could not be assigned because of its weak intensity and poor signal-to-noise ratio.

\section{Conclusions}

A rotationally resolved photoelectron spectrum of the H$_2$O$^+$ $\tilde{\mathrm{A}}^+$ (0,$v_2^{\textrm{lin}}$,0)-H$_2$O $\tilde{\mathrm{X}}$ (0,0,0) transitions with $v_2^{\textrm{lin}}$ between 3 and 8 has been recorded. Because of the low temperature of the supersonic expansion, only ground-state levels with $K_{\mathrm{a}}=0$ were significantly populated, which strongly favoured the observation of $\Sigma$ and $\Pi$ vibronic states of the $\tilde{\mathrm{A}}^+$ state. The transitions that are expected to be dominant according to photoionization selection rules and simple considerations concerning the symmetry of the molecular orbitals, are those with $\Delta K_{\mathrm{a}}=\pm 1$, $\Delta N=0$,$\pm\ 1$ accompanied by the emission of even-\textit{l} photoelectron partial waves. Such transitions were found to be dominant in the spectra of the $\tilde{\mathrm{A}}^+\ \Pi (0,v_2^{\textrm{lin}}=6$ and 8, 0)
$\leftarrow \tilde{\mathrm{X}}(000)$ bands. Below 10 K, only the ground rotational states of para ($0_{00}$) and ortho ($1_{01}$) H$_2$O are significantly populated and both have $K_{\mathrm{a}}=0$. Transitions from these levels to $\tilde{\mathrm{A}}^+$ levels of $\Sigma$ vibronic symmetry are therefore intrinsically $\Delta K_{\mathrm{a}}=0$ transitions and must be accompanied by the emission of odd-\textit{l} photoelectron partial waves. We therefore expect the photoelectron angular distributions associated with transitions to $\tilde{\mathrm{A}}^+$ states of $\Sigma$ symmetry to be strongly temperature dependent and to change from a regime where even-\textit{l} partial waves are favoured at high temperatures to a regime where odd-\textit{l} partial waves are dominant at low temperatures.

The rovibronic term values extracted from our photoelectron spectra of the $\tilde{\mathrm{A}}^+\ \Pi (060)$, $\Sigma (070)$ and $\Pi (080)$ are in good agreement with the term values determined in the analysis of the $\tilde{\mathrm{A}}^+$-$\tilde{\mathrm{X}}^+$ band system \cite{Lew1973, Lew1976, Huet1997}. They are also in good agreement with the term values calculated \textit{ab initio} by Brommer \textit{et al.} \cite{Brommer1992} who, however, optimized their \textit{ab initio} potential to best reproduce the available experimental data. The deviations between calculated and experimental $\tilde{\mathrm{A}}^+$ state term values deteriorates as one moves to lower energies. This behaviour indicates that the $\tilde{\mathrm{A}}^+$ state remains insufficiently characterized at low energies. 
\section{Acknowledgements}
We thank Ugo Jacovella and Urs Hollenstein (both ETH Zurich) for helpful discussions and experimental assistance. We also thank Alexander Alijah (University of Reims Champagne-Ardenne), Geoffrey Duxbury (University of Strathclyde) and Th\'er\`ese Huet (University of Lille 1) for useful correspondence concerning the $\tilde{\mathrm{A}}^+$ state of H$_2$O$^+$. This work is supported financially by the Swiss National Foundation under project Nr. 200020-159848.

\end{document}